\documentclass[prl,twocolumn,superscriptaddress, nofootinbib]{revtex4}

\usepackage{amsfonts}
\usepackage{amsmath}    % for subequations
\usepackage[normalem]{ulem}
\usepackage{bm}
\usepackage{graphicx}
\usepackage{natbib}

\newcommand{\ket}[1]{\vert#1\rangle}

\def\opone{\leavevmode\hbox{\small1\kern-3.8pt\normalsize1}}
\usepackage{color}

\begin{document}

\title{A telecom-wavelength atomic quantum memory in optical fiber for heralded polarization qubits}

\author{Jeongwan Jin}
\altaffiliation{Present address: Institute for Quantum Computing, Department of Physics and Astronomy, University of Waterloo, 200 University Ave West, Waterloo, Ontario, Canada}
\affiliation{Institute for Quantum Science and Technology, and Department of Physics \& Astronomy, University of Calgary, 2500 University Drive NW, Calgary, Alberta T2N 1N4, Canada}
\author{Erhan Saglamyurek\footnote{J.J. and E.S. contributed equally to this work.}}
\affiliation{Institute for Quantum Science and Technology, and Department of Physics \& Astronomy, University of Calgary, 2500 University Drive NW, Calgary, Alberta T2N 1N4, Canada}
\author{Marcel.l\'{i} Grimau Puigibert}
\affiliation{Institute for Quantum Science and Technology, and Department of Physics \& Astronomy, University of Calgary, 2500 University Drive NW, Calgary, Alberta T2N 1N4, Canada}
\author{Varun Verma}
\affiliation{National Institute of Standards and Technology, Boulder, Colorado 80305, USA}
\author{Francesco~Marsili}
\affiliation{Jet Propulsion Laboratory, California Institute of Technology, 4800 Oak Grove Drive, Pasadena, California 91109, USA}
\author{Sae~Woo~Nam}
\affiliation{National Institute of Standards and Technology, Boulder, Colorado 80305, USA}
\author{Daniel~Oblak}
\affiliation{Institute for Quantum Science and Technology, and Department of Physics \& Astronomy, University of Calgary, 2500 University Drive NW, Calgary, Alberta T2N 1N4, Canada}
\author{Wolfgang~Tittel} 
\email{wtittel@ucalgary.ca}
\affiliation{Institute for Quantum Science and Technology, and Department of Physics \& Astronomy, University of Calgary, 2500 University Drive NW, Calgary, Alberta T2N 1N4, Canada}

\begin{abstract}
Photon-based quantum information processing promises new technologies including optical quantum computing, quantum cryptography, and distributed quantum networks. Polarization-encoded photons at telecommunication wavelengths provide a compelling platform for practical realization of these technologies. However, despite important success towards building elementary components compatible with this platform, including sources of entangled photons, efficient single photon detectors, and on-chip quantum circuits, a missing element has been atomic quantum memory that directly allows for reversible mapping of quantum states encoded in the polarization degree of a telecom-wavelength photon. Here we demonstrate the quantum storage and retrieval of polarization states of heralded single-photons at telecom-wavelength by implementing the atomic frequency comb protocol in an ensemble of erbium atoms doped into an optical fiber. Despite remaining limitations in our proof-of-principle demonstration such as small storage efficiency and storage time, our broadband light-matter interface reveals the potential for use in future quantum information processing.
\end{abstract}

\maketitle
Photonic quantum communication technologies rely on encoding quantum information (e.g.~qubits) into single photons, and processing and distributing it to distant locations. In principle, a qubit can be encoded into any photonic degree of freedom, but the polarization degree has often been a preferred choice due to the ease of performing single qubit manipulations and projection measurements using wave-plates and polarizing beam-splitters, and the availability of polarization-entangled photon-pair sources~\cite{kwiat1995a}. On this background many seminal demonstrations of quantum information processing have employed polarization qubits, such as teleportation~\cite{bouwmeester1997a}, entanglement swapping~\cite{pan1998a}, and secure quantum key distribution over more than hundred kilometers~\cite{peng2007a}, pointing towards the possibility to build secure quantum networks~\cite{gisin2002a}. 

In addition to polarization encoding, the realization of future quantum networks will be greatly facilitated by the use of existing fiber optic infrastructure, which allows low propagation loss for photons at wavelengths around 1550 nm, commonly referred to as telecom-wavelengths~\cite{kimble2008a}. Nevertheless, propagation loss still limits the distance of such quantum links to a few hundreds kilometers. This distance barrier may be overcome by the development of quantum repeaters~\cite{sangouard2011a}. These, in turn, rely on the ability to store photonic qubits in quantum memories to overcome the probabilistic nature of photon transmission through lossy quantum channels and emission from currently used single-photon sources~\cite{lvovsky2009a, bussieres2013a}. However, to date, no quantum memory for polarization qubits encoded into telecom wavelength photons has been demonstrated. In fact, until recently \cite{saglamyurek2015a}, direct quantum storage of telecom-wavelength photons was an unsolved problem. The most promising candidate, erbium, has the required transition wavelength but exhibits atomic-level dynamics that have shown to be challenging for non-classical light storage~\cite{lauritzen2010a, dajcgewand2014a}. This has prompted indirect approaches such as quantum state teleportation~\cite{bussieres2014a} or coherent conversion techniques~\cite{maring2014a} from telecom-wavelength photons to photons at other wavelengths at which existing quantum memories can operate.

\begin{figure*}[ht!]
\begin{center}
\includegraphics[width=\textwidth,angle=0]{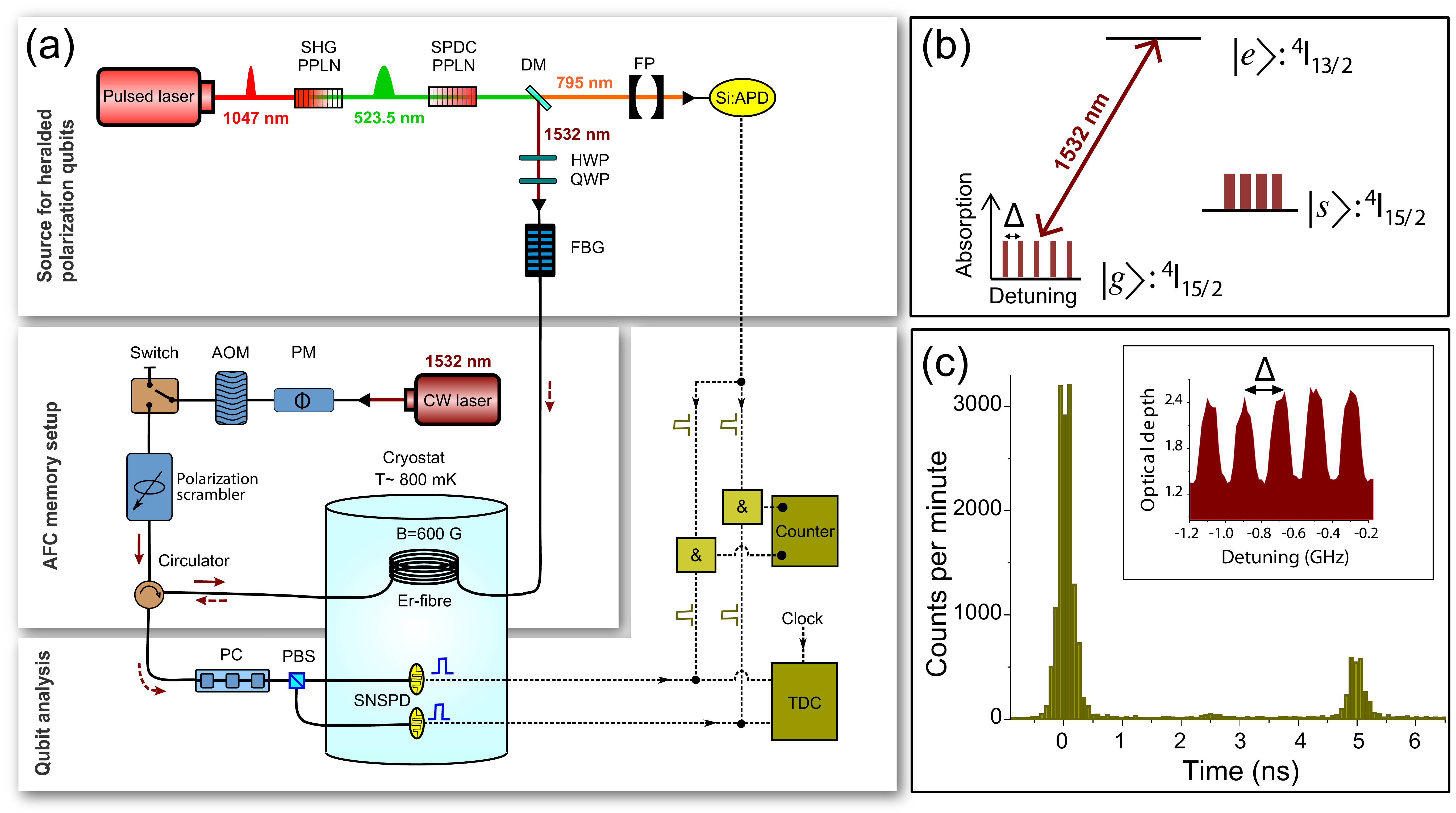}
\caption{\textbf{Experimental setup.} 
\textbf{a}, sketch of setup:
In the heralded polarization qubit source, 6~ps long, 80~MHz repetition-rate pulses at 1047 nm wavelength are frequency doubled and subsequently downconverted in periodically poled lithium niobate crystals (PPLN), thus generating photon pairs at 795~nm and 1532~nm (telecom) wavelengths. Detection of the 795~nm photon by a Si-APD (60\% detection efficiency and 600~ps timing-jitter) heralds the presence of the telecom photon, and a half-wave-plate (HWP) and quarter-wave-plate (QWP) prepare it in any desired polarization qubit state. 
The quantum memory consists of a 20-meter-long erbium doped fiber exposed to a 600~G magnetic field and cooled to 0.8~K using a cryostat. The polarization-insensitive AFC is prepared by optical pumping with continuous wave (CW) laser light at 1532~nm, which is frequency and amplitude modulated by phase modulator (PM) and an acousto-optic modulator (AOM), respectively, and polarization-randomized by a polarization scrambler with 5~KHz cycling frequency.
For the qubit analysis, the recalled photons pass through a fiber-optic polarization controller (PC) and polarizing beam splitter (PBS) after which they impinge on one of two superconducting nano-wire single photon detectors (SNSPD) held at 0.8~K (60\% system efficiency and 350~ps time-jitter). A set of AND gates counts coincidences between the qubit detection signals and the heralding signal, and a time-to-digital (TDC) converter additionally provides time mapping of the single detections of the 1532 nm photons compared to the moment of detection of the heralding photons at 795 nm wavelength. 
\textbf{b}, simplified erbium level scheme: 
The AFC is prepared via frequency-selective population transfer from the $^4I_{15/2}$ electronic ground state ($\ket{\mathrm{g}}$) through the $^4I_{13/2}$ excited state ($\ket{\mathrm{e}}$) into an auxiliary (spin) state ($\ket{\mathrm{s}}$).
\textbf{c}, storage of telecom-wavelength photons in a broadband AFC memory: 
Detection signal from a single SNSPD recorded by the TDC, showing directly transmitted and stored photons spaced by 5~ns. The inset shows a 1~GHz wide section of a typical AFC of 8~GHz total bandwidth. The background absorption in the AFC arises from imperfect optical pumping.}  
\label{setup}
\end{center}
\end{figure*}

Furthermore, storage of polarization qubits at any wavelength is inhibited by the fact that the interaction of polarized light with most atomic media has orientational dependence, and birefringence in solid state materials often complicates matters further. In a few experiments -- using light outside the telecom range -- this has been circumvented by storing each polarization component in two spatially separated parts of a memory that each only interact with a single polarization component \cite{england2012a,clausen2012a,gundogan2012a,zhou2012a}. However, this begs the question if there exists a solid-state material in which absorbers in an ensemble are aligned uniformly so that, in essence, all polarizations of light are equally coupled. Recent advances in the development of quantum memories based on erbium doped glass fiber point exactly in this direction \cite{saglamyurek2015a}. Because glass constitutes an amorphous host, the transition dipole moment of embedded rare-earth ions has no preferred direction, resulting in uniform coupling of an ensemble to all light polarizations. Moreover, the possibility to splice fiber-based memories with standard telecommunication fibers promises simple and low-loss integration with other components in a network.

In this letter, we report the faithful storage of polarization states of heralded telecom-wavelength single-photons in collective excitation of an atomic ensemble by implementing the atomic frequency comb (AFC) protocol in a cryogenically cooled erbium-doped optical fiber. 
We verify the memory's ability to store any polarization state with close to equal probability, and demonstrate its quantum nature by showing that recalled single photon quantum states have a near unity fidelity with the originally prepared states.  
%Furthermore we show that our fiber-based storage device provides a unique way for simple and robust integration into telecom-wavelength quantum technologies relying on polarization encoding.

Our experimental setup, sketched in Fig~\ref{setup}, is composed of three main parts: A fiber-based atomic quantum memory, a source of heralded polarization qubits, and a qubit analyzer.
The first step in our experiment is to prepare the AFC memory using the setup shown in the middle panel of Fig.~\ref{setup}a. In this step an optical pumping beam from a continuous wave laser is frequency and amplitude modulated before entering a 20 meter-long erbium-doped fiber cooled to 0.8~K using a cryostat based on a pulse-tube cooler and an adiabatic demagnetization refrigerator. The spectral profile of the pump light is chosen to be a square modulation with a period $\Delta$. As indicated in Fig.~\ref{setup}b, this results in frequency-selective excitation of the erbium ions on their large inhomogeneously broadened transition line at 1532 nm. The polarization of the pump light can be actively scrambled, resulting in uniform absorption of photons in different polarization states, as further discussed below. We maintain the optical pumping for 500~ms, which leads to atoms with transition frequencies matching the pumping light to accumulate in a long lived electronic Zeeman level that arises under the application of a magnetic field of 600 Gauss. This process allows us to  generate an 8~GHz broad comb shaped absorption profile (a 1~GHz broad section is shown in the inset of Fig.~\ref{setup}c) i.e. an AFC, having a tooth spacing of $\Delta=200$~MHz and finesse (the ratio of peak spacing to width) of 2. To avoid photons stemming from spontaneous decay of atoms in the excited state during qubit recall, a wait time of 300~ms is set after the optical pumping. During 700~ms, following the wait time, we can map a photon onto our AFC, giving rise to collective excitation of the erbium ions in what is commonly referred to as a Dicke state:
\begin{equation}
	\label{eq:afcqstate}
	\left| \Psi  \right\rangle =\frac{1}{\sqrt{N}}\sum _{j=1}^{N} c_{j} e^{i2\pi m_j\Delta t} e^{-ikz_j}  \left| g_{1} ,\cdots e_{j} ,\cdots g_{N}  \right\rangle \ .	
\end{equation}
Here, ${\left| g_{j}  \right\rangle}$ (${\left| e_{j}  \right\rangle}$) is the ground (excited) state of atom $j$, $m_j\Delta$ is the detuning of the transition frequency of the atom from the photon carrier frequency, $z_j$ is the atom's position measured along the propagation direction of the light, and the factor $c_j$ depends on the resonance frequency and position of the atom. Due to the different atomic transition frequencies, each excitation term in the Dicke state accumulates a different phase over time. However, the periodic structure of the transition frequencies of atomic absorbers in the AFC leads the phases to realign at a time given by the inverse of the peak spacing $t_{storage}=1/\Delta$~[\citenum{afzelius2010a}]. Consequently, for $\Delta$=200 MHz, an input photon is retrieved after $t_{storage}=5$~ns in the originally encoded state, as shown in Fig.~\ref{setup}c.

The second step is to prepare heralded polarization qubits, that are compatible with our light-matter interface, using the photon pair source shown in the top panel of Fig.~\ref{setup}a. To this end, short pulses produced from a mode-locked laser operating at 1047~nm wavelength are frequency doubled via second harmonic generation (SHG). The resulting pulses at 523~nm are sent to a periodically poled lithium niobate (PPLN) crystal where they take part in a spontaneous parametric down conversion (SPDC) process. The phase matching condition for SPDC results in a pair of photons at 1532 nm and 795 nm, named \textit{signal} and \textit{idler} photons, respectively. In order to match the acceptance bandwidth of our storage device, the bandwidths of the 1532 nm and 795 nm photons are filtered down to nearly 10 GHz using a Fiber-Bragg grating (FBG) and a Fabry-Perot (FP) cavity, respectively. The 795 nm photons are then directly detected by a Si-APD single-photon detector to provide a heralding signal. The 1532 nm photons pass through a free-space Half-Wave Plate (HWP) and Quarter-Wave Plate (QWP), which allow us to encode the polarization states
\begin{equation}
 \label{projection}
 \ket{\psi}=\cos\theta\ket{H}+e^{i\varphi}\sin\theta\ket{V} \ ,%\hspace{1cm}\alpha^2+\beta^2=1 , 
 \end{equation}  
where $\ket{H}$ and $\ket{V}$ denote a horizontally and vertically polarized 1532~nm photon, and $\cos\theta\equiv\alpha$ and $\sin\theta\equiv\beta$ are the probability amplitudes for $\ket{H}$ and $\ket{V}$, respectively. At this point the prepared polarization qubits are sent to the prepared AFC memory for storage and recall.

The final step is to analyze the recalled qubits using the analyzer shown in the bottom panel of Fig.~\ref{setup}a. Its optical part consists of a polarization controller (PC) followed by a polarizing beam-splitter (PBS) with super-conducting nanowire single-photon detectors (SNSPD) at each output. These allow making projection measurements onto any set of orthogonal polarization qubit states. The electronic part consists of logic gates that enable counting coincidences with the heralding signal.

\begin{figure*}
\begin{center}
\includegraphics[width=1\textwidth,angle=0]{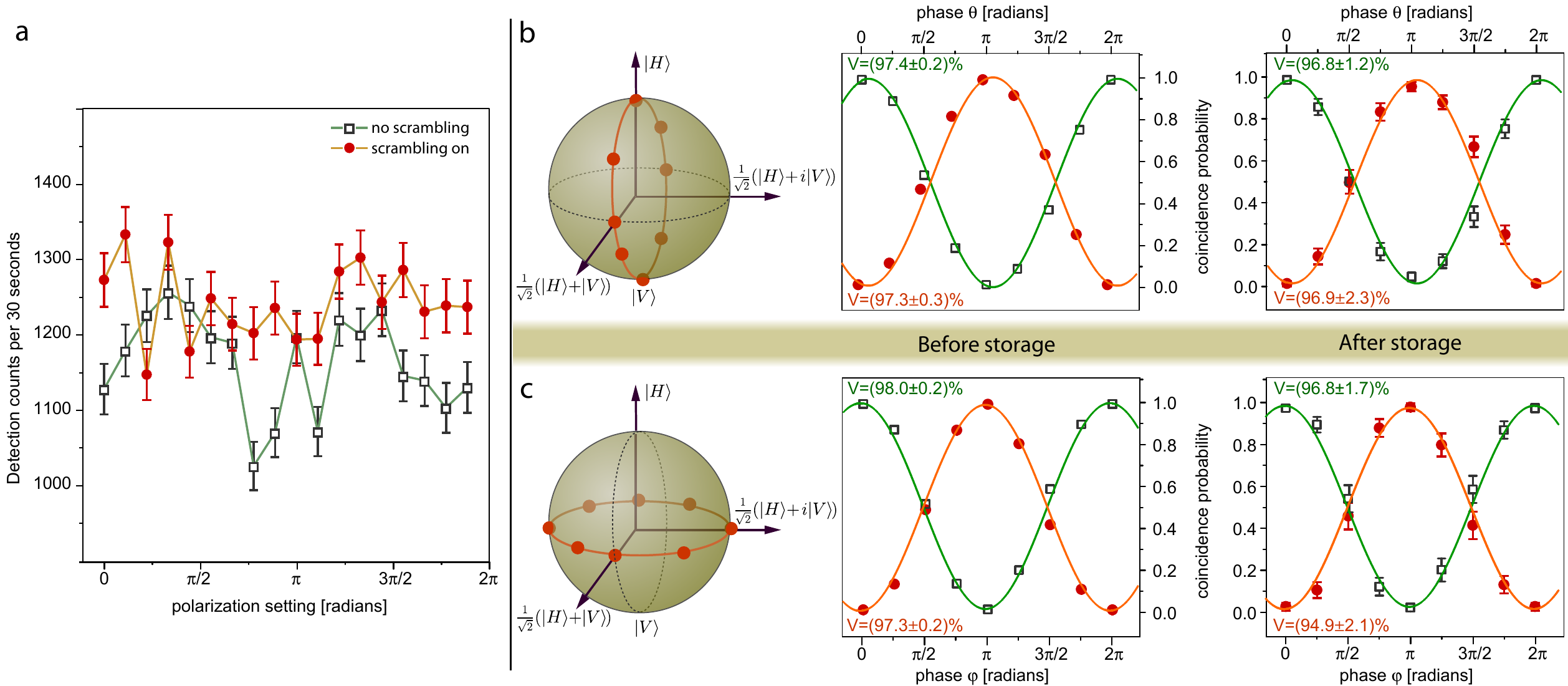}
\caption{\textbf{Storage of polarization states} \textbf{a}, polarization insensitive operation of fiber-memory:
Detection counts of recalled photons as recorded on the TDC (summed over 5 bins, each amounting to 80 ps) as function of input HWP setting and normalized w.r.t. the polarization dependence of the SNSPD efficiency as further explained in the main text.
\textbf{b}, measured coincidence probabilities for qubit states of the form $\ket{\psi}=\cos\theta\ket{H}+\sin\theta\ket{V}$ projected onto $\ket{H}$ (squares) and $\ket{V}$ (circles) before (l.h.s.) and after (r.h.s.) storage in the AFC.
\textbf{c}, 
measured coincidence probabilities for qubit states of the form $\ket{\psi}=\frac{1}{\sqrt{2}}(\ket{H}+e^{i\varphi}\ket{V}$ projected onto $\ket{+}=\frac{1}{\sqrt{2}}(\ket{H}+\ket{V}$ (squares) and $\ket{-}=\frac{1}{\sqrt{2}}(\ket{H}-\ket{V}$ (circles) before (l.h.s.) and after (r.h.s.) storage in the AFC.
%storage of qubits in the state of $\ket{\psi}=\frac{1}{\sqrt{2}}(\ket{H}+e^{i\varphi}\ket{V}$. Each input qubit is prepared by varying $\varphi$ using the HWP and QWP, and projected onto $\ket{\psi}=\frac{1}{\sqrt{2}}(\ket{H}\pm\ket{V}$ after storage in the AFC. The measurement and analysis procedures are the same as described in \textbf{b}.
%
Probabilities are calculated by dividing the coincidence counts at each PBS output during 5 minutes by the total coincidence counts (i.e. the sum of the counts in the two PBS outputs), which is approximately 1 Hz. 
Error bars are based on Poissonian detection statistics and all visibility values are based on cosine fits to the data.}

\label{visibility}
\end{center}
\end{figure*}

In our measurements we first characterize the polarization sensitivity of our fiber-based atomic memory with and without the polarization scrambling. Initially, we set the polarization state of the 1532~nm photons to be $\ket{H}$ just before the memory input.
% using a fiber-based polarization controller (not shown in Fig.~\ref{setup}a), which is calibrated to a free-space polarization controller. 
Next we gradually change the polarization state between horizontal and vertical by rotating the free-space HWP in steps of 15 degrees. For each polarization setting we store the 1532-nm photons in our AFC memory first without, and then with the polarization scrambler engaged. In this test, we disregard the heralding signal from the 795~nm photon detection and bypass the PBS such that we detect the retrieved 1532~nm photons using a single SNSPD, which has nearly 5\% polarization dependence of its detection efficiency~\cite{marsili2013a}.
In order to account for the variation of detector efficiency with polarization we count the 1532~nm photons directly transmitted through the erbium-fiber without preparing the AFC and use this value to weigh the measurements from the storage experiment (note that the transmission of photons through the unprepared fiber is not polarization dependent).
Figure~\ref{visibility}a shows the weighted detection counts of stored photons as a function of the polarization setting without and with the polarization scrambler. For the former case, we observe that the number of counts, and hence the efficiency of the memory, varies by about 25\% with the polarization setting. This is due to the mismatch of the polarization state of the optical pumping light and the 1532~nm single photons, which counter-propagate through the fiber. This phenomenon is also known as ``polarization hole-burning'' in the operation of erbium-doped fiber amplifiers \cite{becker1999}. However, when the polarization of the optical pumping light is scrambled, the generated AFC in any spatial section of the fiber consists of erbium ions that can be excited by arbitrary polarization state of the heralded single photons. In this case, the recall efficiency is near the maximum for all the utilized input polarization states. We attribute the remaining fluctuations in the AFC efficiency (up to 7\%) to the unstable operation of the optical pumping laser over the course of the measurements.
%This shows the polarization insensitive and robust operation of our storage device.

Next, in order to verify the quantum nature of the storage we measure the second-order cross-correlation $g_{si}^{(2)}$ 
 \begin{equation}
 \label{g2}
  g_{si}^{(2)}=\frac{P_{si}}{P_{i}P_{s}} \ ,
	\end{equation}
where $P_{si}$ is the probability of coincidence detection between the heralding signal and the two qubit analyzer outputs combined.  $P_{s}$ and $P_{i}$ are the detection probabilities for the 795~nm photons and the 1532~nm photons, respectively. Measuring a value $g_{si}^{(2)}>2$  indicates the presence of quantum correlations between the photons in each pair, where we assume that the auto-correlation values for the signal and idler photons are in between 1 and 2, corresponding to a coherent and a thermal state, respectively. We find  $g_{si}^{(2)}$ to be $14.1\pm0.3$  before storage and, crucially, $18.4\pm1.2$ after storage, which demonstrates that quantum correlations between the members of the photon pairs are preserved during the storage. 
%The reason for the increase in $g_{si}^{(2)}$ after storage is due to the limited bandwidth of the AFC memory causing some spectral filtering of the photons, which results in a lower effective mean-photon number and thus a larger cross-correlation coefficient.
%
The reason for the increase in $g_{si}^{(2)}$ after storage is due to spectral filtering of the input photons (10~GHz bandwidth) caused by the slightly smaller bandwidth of the AFC memory (8~GHz), which results in a lower effective mean-photon number.

\begin{figure*}[ht!]
\begin{center}
\includegraphics[width=\textwidth,angle=0]{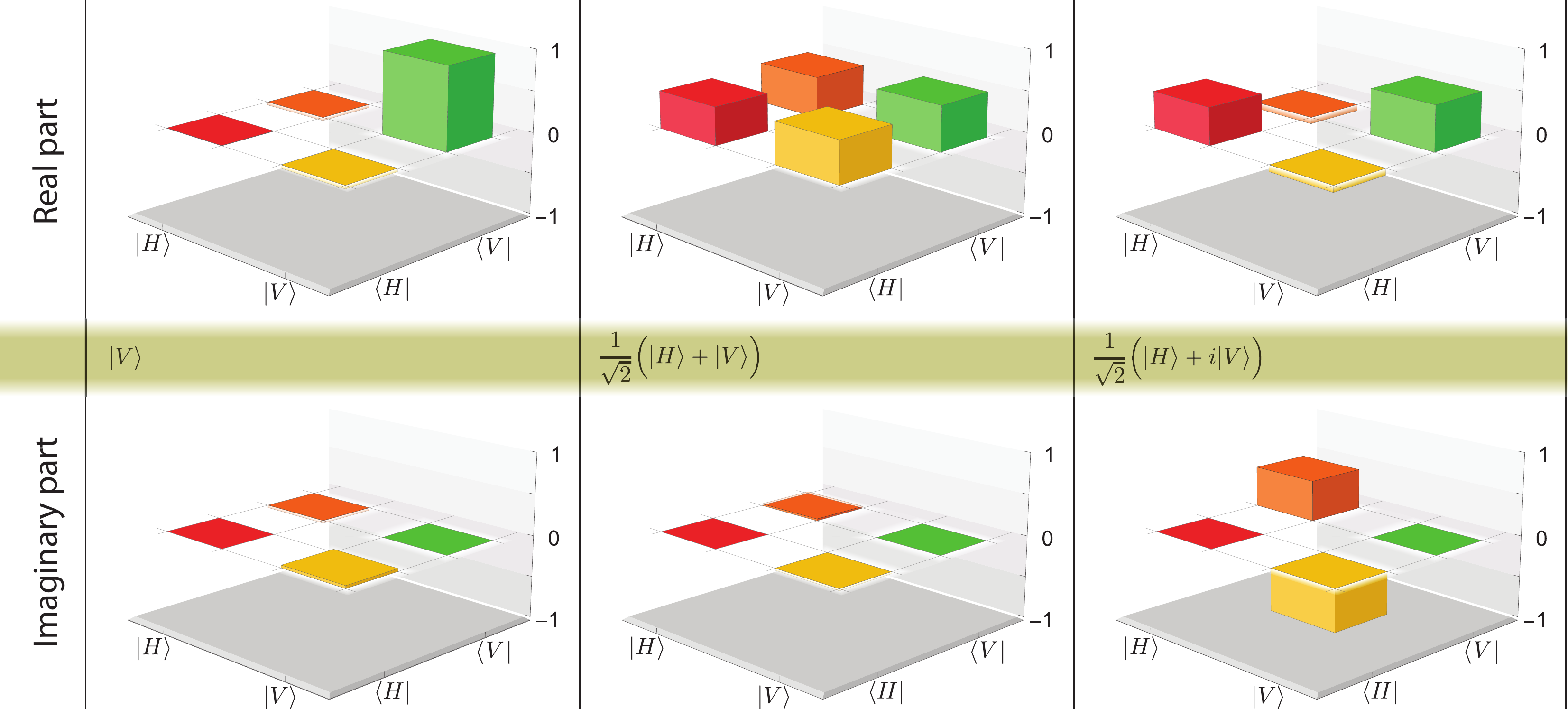}
\caption{\textbf{Reconstructed states of stored qubits.} Density matrices reconstructed from projection measurements of a set of input qubit states after storage of the heralded telecom photon. Fidelities with photons in ideal input states are indicated in Table~\ref{fidelities}.}
\label{tomography}
\end{center}
\end{figure*}

With this in hand, we now demonstrate the storage of heralded polarization qubits in our atomic memory. To begin, we either vary the phase $\theta$ to create the states $\ket{\psi}=\cos\theta\ket{H}+\sin\theta\ket{V}$ and set the qubit analyzer to project onto the states $\ket{H}$ and $\ket{V}$, or we vary $\varphi$ to create the states $\ket{\psi}=(\ket{H} +e^{i\varphi}\ket{V})/\sqrt{2}$ and set the qubit analyzer to project onto $(\ket{H} \pm\ket{V})/\sqrt{2}$. In both cases, the projection rates are expected to vary sinusoidally with the parameter specifying the input state. In order to establish whether our memory introduces any change to the quantum state of the photons, we perform the measurements both when the 1532~nm photons are recalled from the memory, and when they are just passing through the fiber, i.e. when the AFC is not activated. 

For each projection setting we monitor the outputs of the qubit analyzer over 5 minutes and count the detections of the 1532 nm photons that are in coincidence with the detections of the 795 nm photons (heralding signal) using an AND-gate. The data points and the resulting visibility curves in Fig.~\ref{visibility}b-c show the variation of the coincidence probability for the two sets of input states as a function of the input polarization before and after storage. From the fitted visibilities we find the average fidelities $F_{H/V}=(2+\mathcal{V}_H+\mathcal{V}_V)/4$ for projections onto $\ket{H}$ and $\ket{V}$ to be ($98.67 \pm 0.09 $)\% and ($98.4 \pm 0.6$)\% before and after storage, respectively. Similarly for projections onto $(\ket{H} +e^{i\varphi}\ket{V})/\sqrt{2}$ the fidelities are ($98.83 \pm 0.07$)\% and ($97.93 \pm 0.7$)\%. These result show, within experimental uncertainty, that the qubit states have not been altered during storage. 

To confirm this conclusion, we additionally perform quantum state tomography of several qubit states after recall from our AFC memory. In separate measurements we encode the states $\ket{H}$, $\ket{V}$, $(\ket{H}\pm\ket{V})/\sqrt{2}$ and $(\ket{H}\pm i\ket{V})/\sqrt{2}$, which form three mutually unbiased qubit bases, and map them to our memory. 

We retrieve the photons and, for each qubit, set the analyzers to project them onto the states $\ket{H}$ and $\ket{V}$, $(\ket{H}\pm\ket{V})/\sqrt{2}$, and $(\ket{H}\pm i\ket{V})/\sqrt{2}$. Using the outcomes of the coincidence measurements described previously, we reconstruct the quantum state of each retrieved qubit in terms of the density matrix \cite{altepeter2005a}, examples of which are shown in Fig.~\ref{tomography}. The fidelities of the reconstructed density matrices with respect to those of the qubit states we intended to encode are given in Table~\ref{fidelities}. For all states the fidelities are 0.99 or above, which confirms that the polarization qubit states are well preserved during the memory storage and recall. We associate the small difference of these fidelities from unity with imperfect state preparation and thus not as being due to the storage.

\begin{table}
	\begin{tabular}{ c | c }
		%\hline
		\hspace{10pt} Target qubit \hspace{10pt} & \hspace{15pt} Fidelity \hspace{15pt} \\[2pt]
		\hline
		$\ket{H}$ & $(99.7\pm 1.7)\%$ \\[2pt]
	 	$\ket{V}$ & $(99.6\pm 3.0)\%$ \\[2pt]
		$(\ket{H}+\ket{V})/\sqrt{2}$ & $(99.6\pm 2.6)\%$ \\[2pt]
		$(\ket{H}-\ket{V})/\sqrt{2}$ & $(99.5\pm 2.6)\%$ \\[2pt]
		$(\ket{H}+i\ket{V})/\sqrt{2}$ & $(99.0\pm 5.0)\%$ \\[2pt]
		$(\ket{H}-i\ket{V})/\sqrt{2}$ & $(99.7\pm 1.8)\%$ \\
		%\hline
	\end{tabular}
	\caption{Fidelities of reconstructed density matrices with target states.}
	\label{fidelities}
\end{table}

Despite these important results, the performance of our storage device needs to be improved for practical use in quantum photonic applications. First, in our demonstration the recall efficiency is about 1\% (for 5 ns storage), which is is due to imperfect preparation of the AFC (see Fig.~\ref{setup}c). However, as discussed in detail in \cite{saglamyurek2015a}, it is possible to improve the dynamics of the optical pumping significantly with optimal experimental conditions and thus appreciably improve the recall efficiency. Second, the maximum observed storage time in our current implementation is 50~ns. The limitation to the storage time, i.e. coherence time of the optical transition, is mainly due to the coupling of the erbium ions to two-level systems, which are intrinsic to the use of glass-like disordered hosts, as well as magnetic interactions between the erbium ions. Although it is expected that the coherence time in lightly doped erbium fibers, such as ours, exceeds 5~$\mu$s when their temperature approaches 100~mK, it remains a challenge to achieve the storage times required for quantum repeaters. 
Yet, even with short storage times our light-matter interface possesses a number of features -- such as a large time-bandwidth product, multimode storage capacity, and ability to process qubits -- that makes it a good candidate for realizing on-demand single-photon sources~\cite{nunn2013a} or programmable atomic processors~\cite{Saglamyurek2014a}. In addition, erbium-based memories may be employed as quantum interfaces between telecommunication photons and superconducting circuits~\cite{obrien2014a}.

Finally, we note that our implementation provides a pre-programmed delay given by the inverse of the AFC peak spacing. Yet, as we have discussed in \cite{sinclair2014a}, spectrally-multiplexed storage with fixed storage time supplemented by feed-forward controlled recall is sufficient to implement quantum communication schemes. 

In conclusion, we have realized quantum storage of polarization states of light encoded into heralded telecom-wavelength photons by implementing the AFC protocol in an erbium doped optical fiber. Despite current limitations in terms of storage time and efficiency, the large bandwidth and multimode capacity of our light-matter interface
are ideally suited for various applications in a future quantum Internet, e.g. for linear optics quantum computing and photonic quantum state processing. Furthermore, the robust and simple fiber-based platform of our light-matter interface offers full compatibility with quantum photonics relying on telecom-fiber technology.  

We thank Raju Valivarthi, Qiang Zhou, Matthew D. Shaw, Vladimir Kiselyov for useful discussions and technical support, and gratefully acknowledge support through Alberta Innovates Technology Futures (AITF) and the National Science and Engineering Research Council of Canada (NSERC). W.T. is a senior fellow of the Canadian Institute for Advanced Research (CIFAR). V.B.V. and S.W.N. acknowledge partial funding for detector development from the Defense Advanced Research Projects Agency (DARPA) Information in a Photon (InPho) program. Part of the research was carried out at the Jet Propulsion Laboratory, California Institute of Technology, under a contract with the National Aeronautics and Space Administration.

%\vspace{0.5cm}
%\section*{References}

%\bibliography{/Users/danieloblak/Documents/Proffi/litterature/motherbib}

\begin{thebibliography}{10}

\bibitem{kwiat1995a}
P.~G. Kwiat, K. Mattle, H. Weinfurter, A. Zeilinger, A.~V.
  Sergienko, and Y. Shih.
%\newblock New high-intensity source of polarization-entangled photon pairs.
\newblock {\em Phys. Rev. Lett.}, \textbf{75}, 4337--4341 (1995).

\bibitem{bouwmeester1997a}
D. Bouwmeester, J.-W. Pan, K. Mattle, M. Eibl, H. Weinfurter,
  and A. Zeilinger.
%\newblock Experimental quantum teleportation.
\newblock {\em Nature}, \textbf{390}, 575--579 (1997).

\bibitem{pan1998a}
J.-W. Pan, D. Bouwmeester, H. Weinfurter, and A. Zeilinger.
%\newblock Experimental entanglement swapping: Entangling photons that never interacted.
\newblock {\em Phys. Rev. Lett.}, \textbf{80}, 3891--3894 (1998).

\bibitem{peng2007a}
C.-Z. Peng, J. Zhang, D. Yang, W.-B. Gao, H.-X. Ma, H. Yin, H.-P.
  Zeng, T. Yang, X.-B. Wang, and J.-W. Pan.
%\newblock Experimental long-distance decoy-state quantum key distribution based on polarization encoding.
\newblock {\em Phys. Rev. Lett.}, \textbf{98}, 010505 (2007).

\bibitem{gisin2002a}
N. Gisin, G. Ribordy, W. Tittel, and H. Zbinden.
%\newblock Quantum cryptography.
\newblock {\em Rev. Mod. Phys.}, \textbf{74}, 145--195 (2002).

\bibitem{kimble2008a}
H.~J. Kimble.
%\newblock The quantum internet.
\newblock {\em Nature}, \textbf{453}, 1023--1030 (2008).

\bibitem{sangouard2011a}
N. Sangouard, C. Simon, H. de~Riedmatten, and N. Gisin.
%\newblock Quantum repeaters based on atomic ensembles and linear optics.
\newblock {\em Rev. Mod. Phys.}, \textbf{83}, 33--80 (2011).

\bibitem{lvovsky2009a}
A.~I. Lvovsky, B.~C. Sanders, and W. Tittel.
%\newblock Optical quantum memory.
\newblock {\em Nature Photon.}, \textbf{3}, 706--714 (2009).

\bibitem{bussieres2013a}
F. Bussi{\`e}res, N. Sangouard, M. Afzelius, H.
  de~Riedmatten, C. Simon, and W. Tittel.
%\newblock Prospective applications of optical quantum memories.
\newblock {\em J. Mod. Opt.}, \textbf{60}, 1519--1537 (2013).

\bibitem{lauritzen2010a}
B. Lauritzen, J. Min{\'a}{\v r}, H. de~Riedmatten,
  M. Afzelius, N. Sangouard, C. Simon, and N. Gisin.
%\newblock Telecom-wavelength solid-state memory at the single photon level.
\newblock {\em Phys. Rev. Lett.}, \textbf{104}, 080502 (2010).

\bibitem{dajcgewand2014a}
J.~Dajczgewand, J. L. Le~Gou{\"e}t, A.~Louchet-Chauvet, and T.~Chaneli{\`e}re.
%\newblock Large efficiency at telecom wavelength for optical quantum memories.
\newblock {\em Opt. Lett.}, \textbf{39}, 2711--2714 (2014).

\bibitem{bussieres2014a}
F. Bussi{\`e}res \textit{et al.}
%\newblock Quantum teleportation from a telecom-wavelength photon to a solid-state quantum memory.
\newblock {\em Nature Photon.}, \textbf{8}, 775--778 (2014).

\bibitem{maring2014a}
N. Maring, K. Kutluer, J. Cohen, M, Cristiani, M.
  Mazzera, P.~M Ledingham, and H. de~Riedmatten.
%\newblock Storage of up-converted telecom photons in a doped crystal.
\newblock {\em New J. Phys.}, \textbf{16}, 113021 (2014).

\bibitem{england2012a}
D.~G England, P.~S. Michelberger, T.~F.~M. Champion, K.~F. Reim, K.~C. Lee, M.~R.
  Sprague, X.-M. Jin, N.~K. Langford, W.~S. Kolthammer, J. Nunn, and I.~A.
  Walmsley.
%\newblock High-fidelity polarization storage in a gigahertz bandwidth quantum memory.
\newblock {\em J. Phys. B}, \textbf{45}, 24008 (2012).

\bibitem{clausen2012a}
C. Clausen, F. Bussi\`eres, M. Afzelius, and N. Gisin.
%\newblock Quantum storage of heralded polarization qubits in birefringent and anisotropically absorbing materials.
\newblock {\em Phys. Rev. Lett.}, \textbf{108}, 190503 (2012).

\bibitem{gundogan2012a}
M. G\"undo\u{g}an, P.~M. Ledingham, A. Almasi, M. Cristiani, and H. de~Riedmatten.
%\%newblock Quantum storage of a photonic polarization qubit in a solid.
\newblock {\em Phys. Rev. Lett.}, \textbf{108}, 190504 (2012).

\bibitem{zhou2012a}
Z.-Q. Zhou, W.-B. Lin, M. Yang, C.-F. Li, and G.-C. Guo.
%\newblock Realization of reliable solid-state quantum memory for photonic polarization qubit.
\newblock {\em Phys. Rev. Lett.}, \textbf{108}, 190505 (2012).

\bibitem{saglamyurek2015a}
E. Saglamyurek, J. Jin, V.~B. Verma, M.~D. Shaw, F. Marsili, S.~W. Nam, D. Oblak, and W. Tittel.
%\newblock Quantum storage of entangled telecom-wavelength photons in an erbium-doped optical fibre.
\newblock {\em Nature Photon.}, \textbf{9}, 83--87 (2015).

\bibitem{afzelius2010a}
M. Afzelius, C. Simon, H. de Riedmatten, and N. Gisin.
%\newblock Demonstration of atomic frequency comb memory for light with spin-wave storage.
\newblock {\em Phys. Rev. A}, \textbf{79}, 052329 (2009).

\bibitem{marsili2013a}
F.~Marsili, V.~B. Verma, J.~A. Stern, S.~Harrington, A.~E. Lita, T.~Gerrits,
  I.~Vayshenker, B.~Baek, M.~D. Shaw, R.~P. Mirin, and S.~W. Nam.
%\newblock Detecting single infrared photons with 93~$\%$ system efficiency.
\newblock {\em Nature Photon.}, \textbf{7}, 210--214 (2013).

\bibitem{becker1999}
P.~C. Becker, N.~A. Olsson, and J.~R. Simpson.
\newblock {\em Erbium-Doped Fiber Amplifiers - Fundamentals and Technology}.
\newblock Academic Press, 1999.

\bibitem{altepeter2005a}
J.~B. Altepeter, E.~R. Jeffrey, and P.~G. Kwiat.
%\newblock Photonic state tomography.
\newblock {\em Adv. At., Mol., Opt. Phys.}, \textbf{52}, 105--159 (2005).

\bibitem{nunn2013a}
J.~Nunn, N.~K. Langford, W.~S. Kolthammer, T.~F.~M. Champion, M.~R. Sprague,
  P.~S. Michelberger, X.-M. Jin, D.~G. England, and I.~A. Walmsley.
%\newblock Enhancing multiphoton rates with quantum memories.
\newblock {\em Phys. Rev. Lett.}, \textbf{110}, 133601 (2013).

\bibitem{Saglamyurek2014a}
E.~Saglamyurek, N.~Sinclair, J.~A. Slater, K.~Heshami, D.~Oblak, and W.~Tittel.
%\newblock An integrated processor for photonic quantum states using a broadband light-matter interface.
\newblock {\em New J. Phys.}, \textbf{16}, 065019 (2014).

\bibitem{obrien2014a}
C.~O'Brien, N.~Lauk, S.~Blum, G. Morigi, and M.~Fleischhauer.
%\newblock Interfacing superconducting qubits and telecom photons via a rare-earth doped crystal.
\newblock {\em Phy., Rev. Lett.}, \textbf{113}, 063603 (2014).

\bibitem{sinclair2014a}
N. Sinclair, E. Saglamyurek, H. Mallahzadeh, J.~A. Slater, M. George, R. Ricken, M.~P. Hedges, D. Oblak, C. Simon,
  W. Sohler, and W. Tittel.
%\newblock Spectral multiplexing for scalable quantum photonics using an atomic frequency comb quantum memory and feed-forward control.
\newblock {\em Phys. Rev. Lett.}, \textbf{113}, 053603 (2014).

\end{thebibliography}
%%\bibliographystyle{apsrev4-1}
%\bibliographystyle{unsrt}
%\end{document}

\end{document}